\documentclass{article}

\usepackage{graphicx}
\usepackage{braket}
\usepackage{subfig}

\usepackage{amsmath}
\usepackage{amssymb}

\begin{document}

\title{Non-adiabatic Quantum Wavepacket Dynamics Simulation Based on Electronic Structure Calculations using the Variational Quantum Eigensolver}



\author{Hirotoshi Hirai\thanks{e-mail: hirotoshih@mosk.tytlabs.co.jp}\\
\\
\textit{Toyota Central Research and Development Labs., Inc.,}\\
\textit{41-1, Yokomichi, Nagakute, Aichi 480-1192, Japan}\\
\\
\\
Sho Koh\\
\\
\textit{QunaSys Inc., Aqua Hakusan Building 9F,}\\
\textit{1-13-7 Hakusan, Bunkyo, Tokyo 113-0001, Japan}}

\date{\today}
\maketitle

\begin{abstract}
A non-adiabatic nuclear wavepacket dynamics simulation of the H$_2$O$^+$ de-excitation process is performed based on electronic structure calculations using the variational quantum eigensolver.
The adiabatic potential energy surfaces and non-adiabatic coupling vectors are computed with algorithms for noisy intermediate-scale quantum devices, and time propagation is simulated with conventional methods for classical computers.
The results of non-adiabatic transition dynamics from the $\tilde{B}$ state to $\tilde{A}$ state reproduce the trend reported in previous studies, which 
suggests that this quantum-classical hybrid scheme may be a useful application for noisy intermediate-scale quantum devices.
\end{abstract}

\newpage

\section{Introduction}
Elucidation of the non-adiabatic processes \cite{baer2002introduction, worth2004beyond} that appear in photochemical reactions is important to understand the function of photocatalysts \cite{serpone2000photocatalysis} and to determine a visual mechanism \cite{svaetichin1958retinal}.
The quantum wavepacket simulation is a useful tool to investigate the mechanism of non-adiabatic processes.
In the adiabatic representation, it is necessary to compute the adiabatic potential energy surfaces (PESs) and the non-adiabatic couplings (NACs) between each pair of states for nuclear wavepacket simulations. 
Obtaining these physical quantities in high accuracy in vicinity of the intersection (degenerate) region, where the system has strong non-adiabaticity, requires the use of a high-level quantum chemistry method, such as the full configuration interaction (FCI) method. 
However, the computational cost increases exponentially with the size of the system on conventional classical computers, so that application is limited to small molecular systems~\cite{olsen1990passing}.

Quantum computers are considered to be a powerful tool for quantum chemistry computation because the computational cost can be suppressed into the polynomial scale~\cite{mcardle2020quantum}.
For a noise-less quantum computer, a so-called fault tolerant quantum computer (FTQC), the quantum phase estimation algorithm (QPE) \cite{kitaev1995quantum, Cleve1998} has been proposed to compute the eigenenergies of electronic states. 
Aside from the algorithms based on the adiabatic representation, an algorithm has been proposed to directly solve the time evolution of the wavefunction of nuclei and electrons\cite{kassal2008polynomial}.
However, the FTQC requires a sufficient number of qubits for error correction; therefore, it has been estimated that it will take 20 to 30 years to realize this~\cite{langione2019will}. 

Against FTQC, the variational eigenvalue solver (VQE) algorithm \cite{Peruzzo2014, kandala2017hardware} has been reported in recent years to utilize current noisy intermediate-scale quantum (NISQ) devices \cite{preskill2018quantum} that consist of a small number of noisy qubits and allow only shallow quantum circuits.
The VQE algorithm, which was originally focused on the electronic ground state, has been expanded to compute the energies of electronically excited states \cite{nakanishi2019subspace, mcvqe2019, vqd1, vqd2, qse, qeom}.
In addition, an algorithm used to compute NACs has also been proposed \cite{tamiya2020calculating}, so that the components for quantum wavepacket simulations, PESs and NACs, can be obtained using the VQE.

In the present study, the possibility that the NISQ devices can be applied to quantum wavepacket simulations is investigated by non-adiabatic quantum dynamics simulation in the adiabatic representation using a quantum-classical hybrid scheme.
The adiabatic PESs and NACs required in the simulations are computed with the algorithms for NISQ devices using the simulator, and a non-adiabatic quantum dynamics simulation is performed with a conventional algorithm for classical computers \cite{worth2004beyond}.
We demonstrate the relaxation process of H$_2$O$^+$ from $\tilde{B}^2B_2$ (the second excited state) to $\tilde{A}^2A_1$ (the first excited state) can reproduce the trend reported in the previous studies. 

\section{Method}

\subsection{Hamiltonian in the adiabatic representation}
The degrees of freedom of electrons and nuclei must be treated quantum mechanically to accurately describe the dynamics of non-adiabatic systems.
The Hamiltonian of the nuclear-electron system under non-relativity can be written as 
\begin{equation}
H(\{\vec{R}\}, \{\vec{r}\}) = -\sum_{i=1}^N\frac{\hbar^2}{2M_i}\frac{\partial^2}{\partial \vec{R}_i^2}  -\sum_{j=1}^n\frac{\hbar^2}{2m_j}\frac{\partial^2}{\partial \vec{r}_j^2} + V(\{\vec{R}\}, \{\vec{r}\}),
\end{equation}
where $M_i$ and $m_j$ represent the mass of each nucleus and each electron, and $N$ and $n$ represent the numbers of each particle, respectively. 
$\{\vec{R}\} =\{\vec{R}_1, \vec{R}_2, \cdots \vec{R}_N \}$ and $\{\vec{r}\}=\{\vec{r}_1, \vec{r}_2, \cdots \vec{r}_n \}$ are the collective representations for the positional coordinates of each nucleus and each electron, respectively.
$V(\{\vec{R}\}, \{\vec{r}\})$ contains coulombic interactions between electrons, between nuclei, and between electrons and nuclei. 
The dynamics of the quantum system are described by the following time-dependent Schr\"{o}dinger equation,
\begin{equation}
i\hbar \frac{\partial}{\partial t}\Theta(\{\vec{R}\}, \{\vec{r}\}, t) = H \Theta(\{\vec{R}\}, \{\vec{r}\}, t),
\end{equation}
where  the total wavefunction $\Theta(\{\vec{R}\}, \{\vec{r}\}, t)$ is expanded with the eigenstates of the adiabatic electron Hamiltonian, which is defined later, 
\begin{equation}
\Theta(\{\vec{R}\}, \{\vec{r}\}, t) = \sum_l\chi_l(\{\vec{R}\}, t) \psi_l( \{\vec{r}\};  \{\vec{R}\}).
\end{equation}
Such expansion is possible because $\psi_l(\{\vec{r}\}; \{\vec{R}\})$ is a complete set for the system.
Here, $\psi_l(\{\vec{r}\}; \{\vec{R}\})$ is the eigenstates of the adiabatic electron Hamiltonian,
\begin{equation}
H_{el}(\{\vec{r}\}; \{\vec{R}\}) \psi_l(\{\vec{r}\}; \{\vec{R}\})= E_l(\{\vec{R}\})\psi_l(\{\vec{r}\}; \{\vec{R}\})
\end{equation}
where $H_{el}(\{\vec{r}\}; \{\vec{R}\})$ is defined as
\begin{equation}
H_{el}(\{\vec{r}\}; \{\vec{R}\}) = -\sum_{j=1}^n\frac{\hbar^2}{2m_j}\frac{\partial^2}{\partial \vec{r}_j^2} + V(\{\vec{r}\}; \{\vec{R}\}).
\end{equation}
It should be noted that the adiabatic electron Hamiltonian $H_{el}(\{\vec{r}\}; \{\vec{R}\})$ and its eigenvalues and eigenstates are parametrically dependent on $\{\vec{R}\}$.
Here, the expansion coefficient $\chi_l(\{\vec{R}\}, t)$ can be interpreted as a nuclear wavefunction that represents the chemical reactions, and $E_l(\{\vec{R}\})$ gives the PES of the $l$-th electronic excited state.

Using these notations, the previous time-dependent Schr\"{o}dinger equation can be rewritten as 
\begin{equation}
\begin{split}
i\hbar \frac{\partial}{\partial t} \chi_p(\{\vec{R}\}, t) &=[-\sum_i^N\frac{\hbar^2}{2M_i}\frac{\partial^2}{\partial \vec{R}_i^2}+ E_p(\{\vec{R}\})]\chi_p(\{\vec{R}\},t)\\
&-\sum_q\sum_i\frac{\hbar^2}{M_i} \bra{ \psi_p}\frac{\partial}{\partial \vec{R}_i}\ket{\psi_q} \frac{\partial}{\partial \vec{R}_i}\chi_q(\{\vec{R}\},t)\\
&-\sum_q\sum_i\frac{\hbar^2}{2M_i} \bra{ \psi_p}\frac{\partial^2}{\partial \vec{R}_i^2}\ket{\psi_q} \chi_q(\{\vec{R}\},t),
\end{split}
\end{equation}
where $\bra{\psi_p}\frac{\partial}{\partial \vec{R}_i}\ket{\psi_q}$ is the first-order NAC, and this term is now referred to as the NAC vector.
$\bra{\psi_p}\frac{\partial^2}{\partial \vec{R}_i^2}\ket{\psi_q}$ is the second-order NAC coefficient. 
The diagonal components of the second-order NACs give non-adiabatic corrections to the adiabatic PESs. 
On the other hand, the off-diagonal components of the NAC vectors and second-order NAC coefficients have the effect of mixing the adiabatic states.
It should be noted that the first-order NAC term is proportional to the inner product of the NAC vector and the velocity of the nuclear wavepacket, and the non-adiabatic transitions occur when the both are large.
If these non-adiabatic terms are negligibly small, then the time-dependent Schr\"{o}dinger equation under the Born-Oppenheimer approximation is obtained as
\begin{equation}
i\hbar \frac{\partial}{\partial t} \chi_p(\{\vec{R}\}, t)=[-\sum_i^N\frac{\hbar^2}{2M_i}\frac{\partial^2}{\partial \vec{R}_i^2}+ E_p(\{\vec{R}\})]\chi_p(\{\vec{R}\},t).
\end{equation}
This equation shows that the motions of nuclei and electrons can be separated, and the wavepacket $\chi_p(\{\vec{R}\}, t)$ moves on only one adiabatic PES, $E_p(\{\vec{R}\})$. 
This approximation assumes that the non-adiabatic terms are negligibly small because the nuclear mass $\{M_i\}$ is greater than the electron mass, $m$. 
This approximation holds well near the stable structure of the ground state and is the starting point for the theory of first-principles calculations and quantum chemistry calculations for various materials and molecules. 
However, this approximation breaks down in various phenomena involving excited states. 
This can be understood as follows.
The NAC vectors can be rewritten as
\begin{equation}
\bra{\psi_p}\frac{\partial}{\partial \vec{R}_i}\ket{\psi_q}=\frac{\bra{\psi_p}\frac{\partial H_{el}}{\partial \vec{R}_i}\ket{\psi_q}}{E_q-E_p},
\label{eq_nacvec}
\end{equation}
which shows that the NAC vector is inversely proportional to the gap between adiabatic potential energies. 
The adiabatic approximation (Born-Oppenheimer approximation) thus breaks down near the conical intersection where the adiabatic potentials become close to each other. 
For this reason, the NAC vectors cannot be ignored in order to describe the non-adiabatic transition, in which the order of electronic states changes, in the de-excitation process, as often is the case in photochemical reactions. 
Against the first-order NAC, the contribution of the second-order NAC can be considered negligible because the contribution of the NAC vector dominates in the semiclassical limit, $\hbar \rightarrow 0$ \cite{baer2002introduction, worth2004beyond}.
Thereby the second-order NAC is ignored in this study.

\subsection{H$_2$O$^+$ molecule (symmetric extension model, light-heavy-light approximation)}
In this study, the relaxation process of the H$_2$O$^+$ molecule from $\tilde{B}^2B_2$ (the second excited state) to $\tilde{A}^2A_1$ (the first excited state) is studied to demonstrate the proposed method.
The process of ionization of water molecules and subsequent decomposition (generation of H$^+$, OH$^+$, O$^+$, etc.) is important for investigation of the universe (interstellar medium, molecular clouds, planetary atmosphere, in comets) and radiation damage (radiation therapy for cancer) in living organisms \cite{suarez2015nonadiabatic}. 
Therefore, it has been investigated from both experimental and theoretical perspectives \cite{tan1978absolute, werner19953d, gobet2001total, luna2005fragmentation, luna2007water}, and is suitable as a subject for verification of the proposed method. 

There are three nuclei in H$_2$O$^+$, and the Hamiltonian of a three-body system is given in the literature \cite{johnson1986adiabatic} for the case of H$_2$O, 
where the degrees of freedom for rotation and the coupling between rotation and vibration (Coriolis force) are ignored, and only the intramolecular vibration is focused on. 

For further simplification, we assume a symmetric extension model, $r_1 = r_2 = r$, and use a light-heavy-light (LHL) approximation (i.e., $M_O \rightarrow \infty$ from $M_H \ll M_O$). 
The kinetic energy term of the Hamiltonian for H$_2$O$^+$ can be written as follows. 
\begin{equation}
T = -\frac{1}{M_H}\frac{\partial^2}{\partial r^2} - \frac{1}{M_Hr^2}\frac{\partial^2}{\partial \theta^2} -\frac{1}{4M_Hr^2}(1+csc^2(\theta)),
\end{equation}
where $r = r_1 = r_2$ represents the length of OH, $\theta$ represents the angle of H-O-H, and $csc(\theta) = 1/sin(\theta)$. 
The third term is small (0.0005 Ha or less) in the range dealt with in this study ($\theta > 90^{\circ}$) and is thus ignored in this work. 
The above approximations are justified in the relaxation process immediately after excitation and before energy transfer to other vibrational modes occurs \cite{kroes1994photodissociation}. 

The transition to $\tilde{a}^4B_1$ state is also known to occur, either from $\tilde{B}^2B_2$ or $\tilde{A}^2A_1$ states by spin-orbit couplings.
The previous theoretical study \cite{suarez2015nonadiabatic} and experimental result \cite{brundle1968high} indicated that this had a minor effect on the de-excitation process due to the small magnitude of the spin-orbit coupling term.
Therefore, the effect of the spin-orbit coupling is omitted in this study.

\subsection{Computation of PESs and NAC vectors}
\begin{figure}[ht]
\centering
\includegraphics[width=8.6cm]{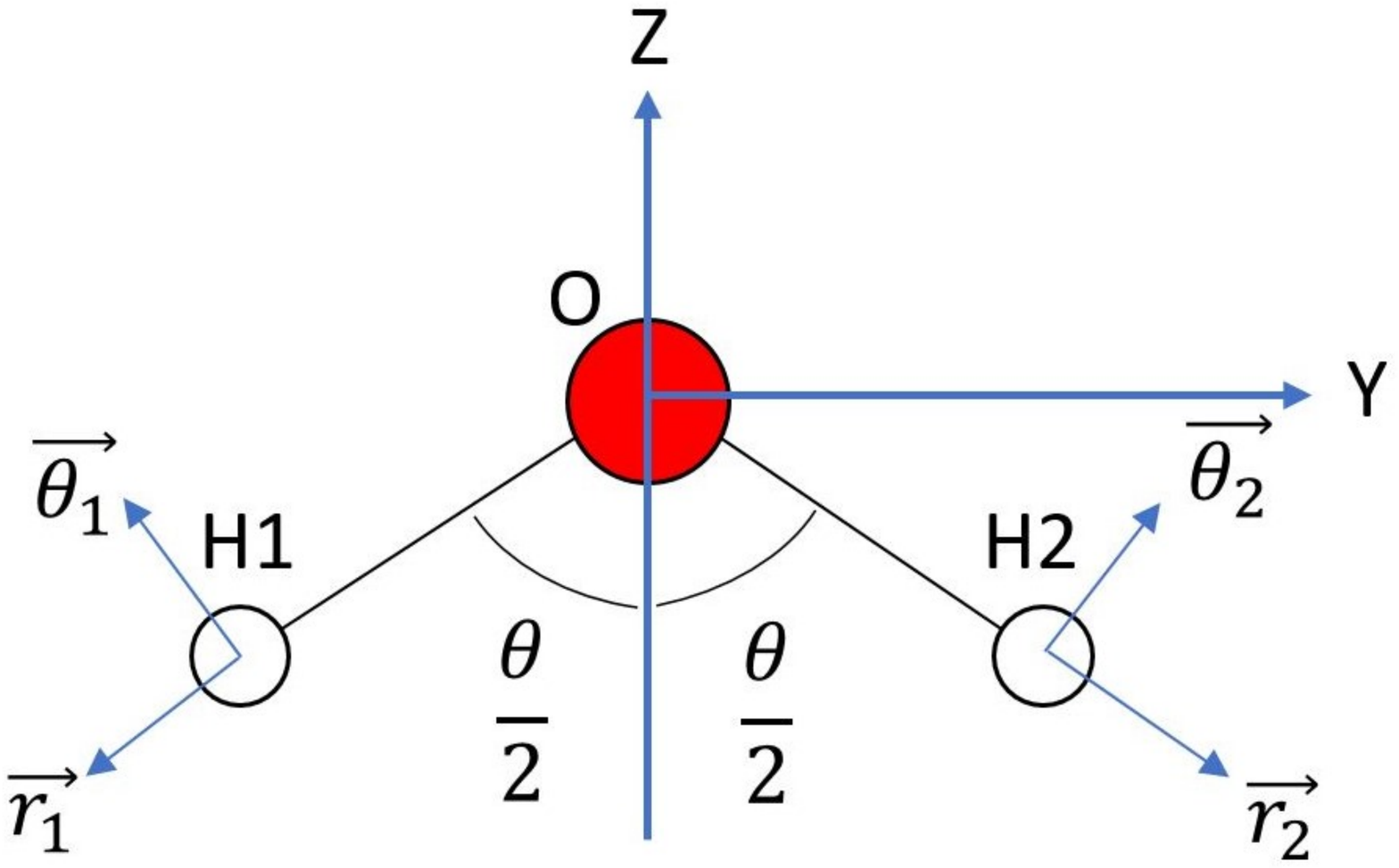}
\caption{Coordinates for adiabatic PESs and NACs computations. The H$_2$O$^+$ molecule is set on the y-z plane where the O atom is placed on the origin and the C$_{2v}$ axis is set along the z-axis.}
\label{fig0}
\end{figure}

The H$_2$O$^+$ molecule is set on the y-z plane where the O atom is placed on the origin and the C$_{2v}$ axis is set along the z-axis, as shown in FIG. \ref{fig0}.
The three lowest electronic states of $\tilde{X}$, $\tilde{A}$ and $\tilde{B}$ are calculated using the subspace-search variational quantum eigensolver (SSVQE) method~\cite{nakanishi2019subspace}, which is an expansion of VQE for obtaining excited states, with the quantum algorithm simulator Qulacs \cite{suzuki2020qulacs} and OpenFermion~\cite{McClean_2020}.
The three minimum molecular orbitals $1b_2$, $3a_1$ and $1b_1$ are taken into the active space to describe the conical intersection structure between the $\tilde{A}$ and $\tilde{B}$ states.
The molecular orbitals used for the second quantized Hamiltonian are optimized using the complete active space self-consistent field (CASSCF) method with the basis set of 6-31G(\textit{d,p}) implemented in the PySCF program \cite{pyscf2020}.

\begin{figure}[ht]
\centering
\includegraphics[width=6.5cm]{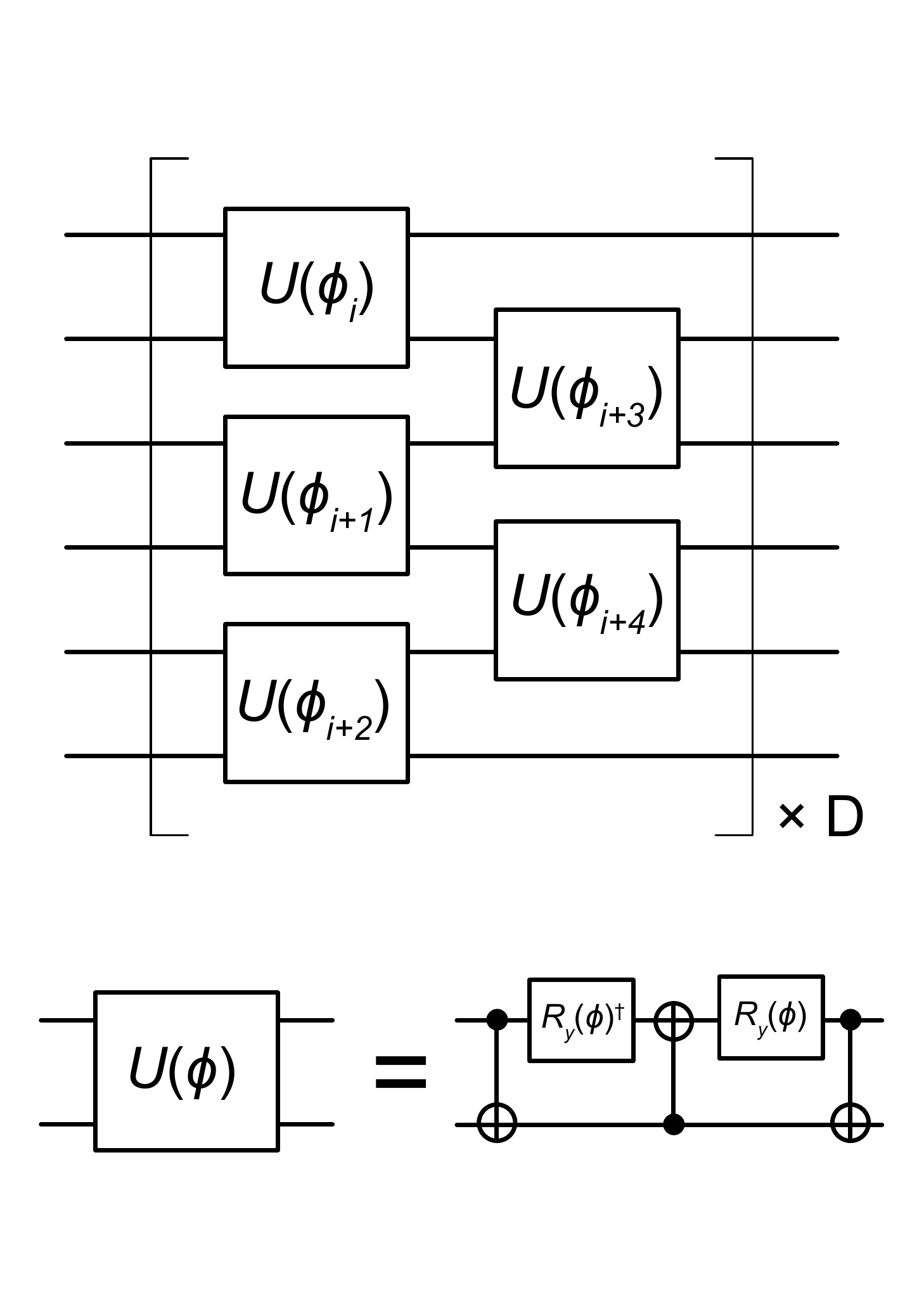}
\caption{Quantum circuit of the symmetry preserving ansatz. $R_y(\phi) = \exp(-i \phi Y/2)$, denotes the y-axis rotation gate and D $= 5$ denotes the depth of the circuit. The rotation angles implemented in two-qubit unitary gates $U(\phi)$ are optimized in VQE calculation.}
\label{fig_spr}
\end{figure}

\begin{figure}[ht]
\centering
\includegraphics[width=8.6cm]{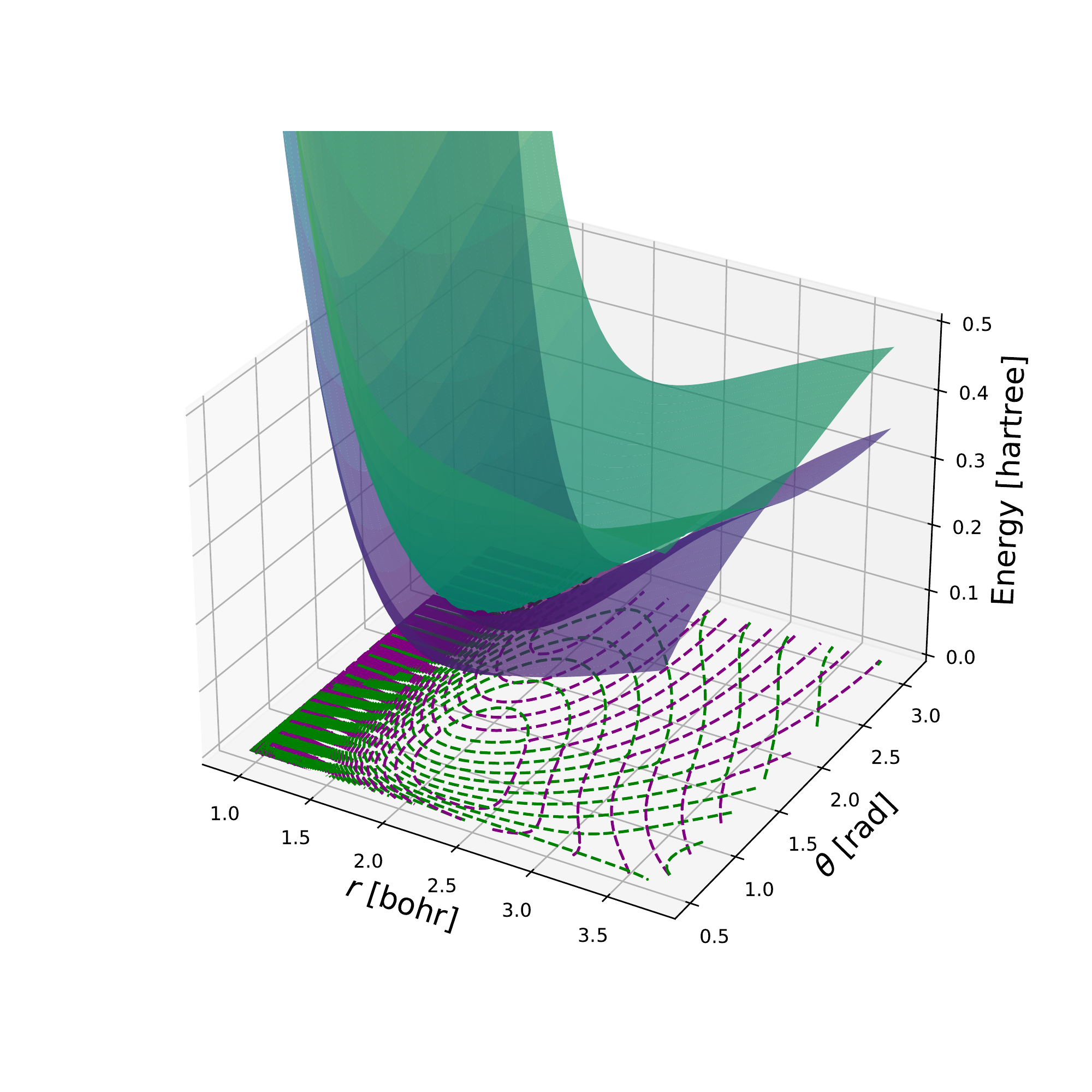}
\caption{Adiabatic PESs of the $\tilde{B}^2B_2$ (green) and $\tilde{A}^2A_1$ (violet) states for the H$_2$O$^+$ molecule. The corresponding contour plots are shown at bottom panel of the figure.}
\label{fig1}
\end{figure}

We use Jordan-Wigner mapping~\cite{aspuru2005simulated} to map spin orbitals onto the qubits so that different spins alternate on 6 qubits.
In the SSVQE calculation, the three initial states of $\ket{101111}$, $\ket{111011}$ and $\ket{1111110}$ on quantum devices are prepared to correspond to the electron configurations with one hole in $1b_1$, $3a_1$ and $1b_1$, respectively.
Symmetry preserving real ansatz \cite{ibe2020calculating} $U(\phi)$ is used to represent the electronic eigenstates of the second quantized Hamiltonian as shown in FIG. \ref{fig_spr}. The weighted energy summation of the three electronic states are given by
\begin{equation}
    E_{SSVQE}(\phi) = w_0 E_0(\phi) + w_1 E_1(\phi) + w_2 E_2(\phi),
\end{equation}
where the weights for the electronic ground state, the first excited state and the second excited state are given as $w_0=9.0, w_1=4.0,$ and $w_2=1.0$, respectively, and the eigenenergies of each electronic state are calculated as 
\begin{eqnarray}
    E_0(\phi) &=& \bra{101111} U^\dagger(\phi) H_{el} U(\phi) \ket{101111}, \\
    E_1(\phi) &=& \bra{111011} U^\dagger(\phi) H_{el} U(\phi) \ket{111011}, \\
    E_2(\phi) &=& \bra{111110} U^\dagger(\phi) H_{el} U(\phi) \ket{111110}.
\end{eqnarray}
The parameters $\phi$ are optimized using the BFGS optimizer in the SSVQE calculation. 
Since only the concept of the method is presented in this study, we do not consider the effect of shot noise.
The obtained adiabatic PESs of the $\tilde{B}^2B_2$ and $\tilde{A}^2A_1$ states are shown in FIG. \ref{fig1}.

After optimization of the electronic states, we compute the NAC matrix elements given in Eq. (\ref{eq_nacvec}).
The operators $\frac{\partial H_{el}}{\partial \vec{R}}$ are calculated using the finite numerical derivative method according to
\begin{equation}
    \frac{\partial H_{el}(\vec{R}_i)}{\partial \vec{R}_i} = \frac{H_{el}(\vec{R}_i + \Delta R \vec{e_i}) - H_{el}(\vec{R}_i - \Delta R \vec{e_i})}{2 \Delta R},
\end{equation}
where $\Delta R$ is set to be 0.001 bohr and $\vec{e_i}$ is the unit vector along $\vec{R}_i$.
The transition amplitude $\bra{\psi_p} \frac{\partial H_{el}(R)}{\partial \vec{R}} \ket{\psi_q}$ can be evaluated on quantum computers, so that the NAC vectors can be computed by dividing this value by the energy gap, $E_q-E_p$, as reported in the previous study \cite{tamiya2020calculating}.

The conversion from the $Y$ and $Z$ NAC vector components to the $r$ and $\theta$ components can be calculated as follows: 
\begin{equation}
\bra{\Psi_p}\frac{\partial}{\partial r_1}\ket{\Psi_q}=-\bra{\Psi_p}\frac{\partial}{\partial Y_1}\ket{\Psi_q}\sin\frac{\theta}{2}-\bra{\Psi_p}\frac{\partial}{\partial Z_1}\ket{\Psi_q}\cos\frac{\theta}{2},
\end{equation}
\begin{equation}
\bra{\Psi_p}\frac{\partial}{\partial r_2}\ket{\Psi_q}=\bra{\Psi_p}\frac{\partial}{\partial Y_2}\ket{\Psi_q}\sin\frac{\theta}{2}-\bra{\Psi_p}\frac{\partial}{\partial Z_2}\ket{\Psi_q}\cos\frac{\theta}{2},
\end{equation}
\begin{equation}
\bra{\Psi_p}\frac{\partial}{\partial \theta_1}\ket{\Psi_q}=-\bra{\Psi_p}\frac{\partial}{\partial Y_1}\ket{\Psi_q}\cos\frac{\theta}{2}+\bra{\Psi_p}\frac{\partial}{\partial Z_1}\ket{\Psi_q}\sin\frac{\theta}{2},
\end{equation}
\begin{equation}
\bra{\Psi_p}\frac{\partial}{\partial \theta_2}\ket\Psi_{q}=\bra{\Psi_p}\frac{\partial}{\partial Y_2}\ket{\Psi_q}\cos\frac{\theta}{2}+\bra{\Psi_p}\frac{\partial}{\partial Z_2}\ket{\Psi_q}\sin\frac{\theta}{2}.
\end{equation}
And the following relationships hold:
\begin{equation}
\bra{\Psi_p}\frac{\partial}{\partial r}\ket{\Psi_q}=\bra{\Psi_p}\frac{\partial}{\partial r_1}\ket{\Psi_q}=\bra{\Psi_p}\frac{\partial}{\partial r_2}\ket{\Psi_q},
\end{equation}
\begin{equation}
\bra{\Psi_p}\frac{\partial}{\partial  \theta}\ket{\Psi_q}=\bra{\Psi_p}\frac{\partial}{\partial \theta_1}\ket{\Psi_q}=\bra{\Psi_p}\frac{\partial}{\partial \theta_2}\ket{\Psi_q}.
\end{equation}

\begin{figure}[ht]
\centering
\subfloat[NAC vector for the $r$-direction]{\includegraphics[width=6.5cm]{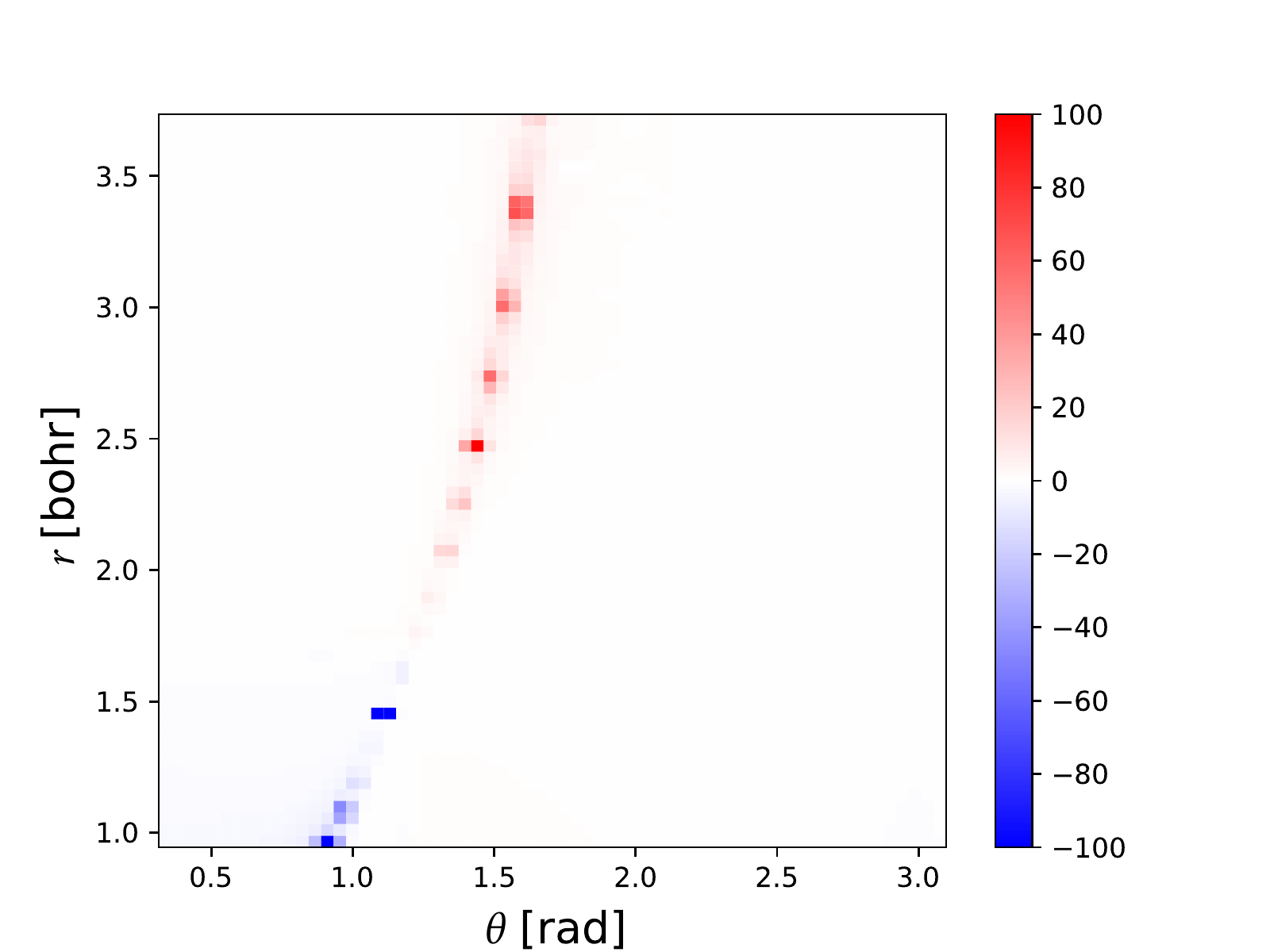}}
\subfloat[NAC vector for the $\theta$-direction]{\includegraphics[width=6.5cm]{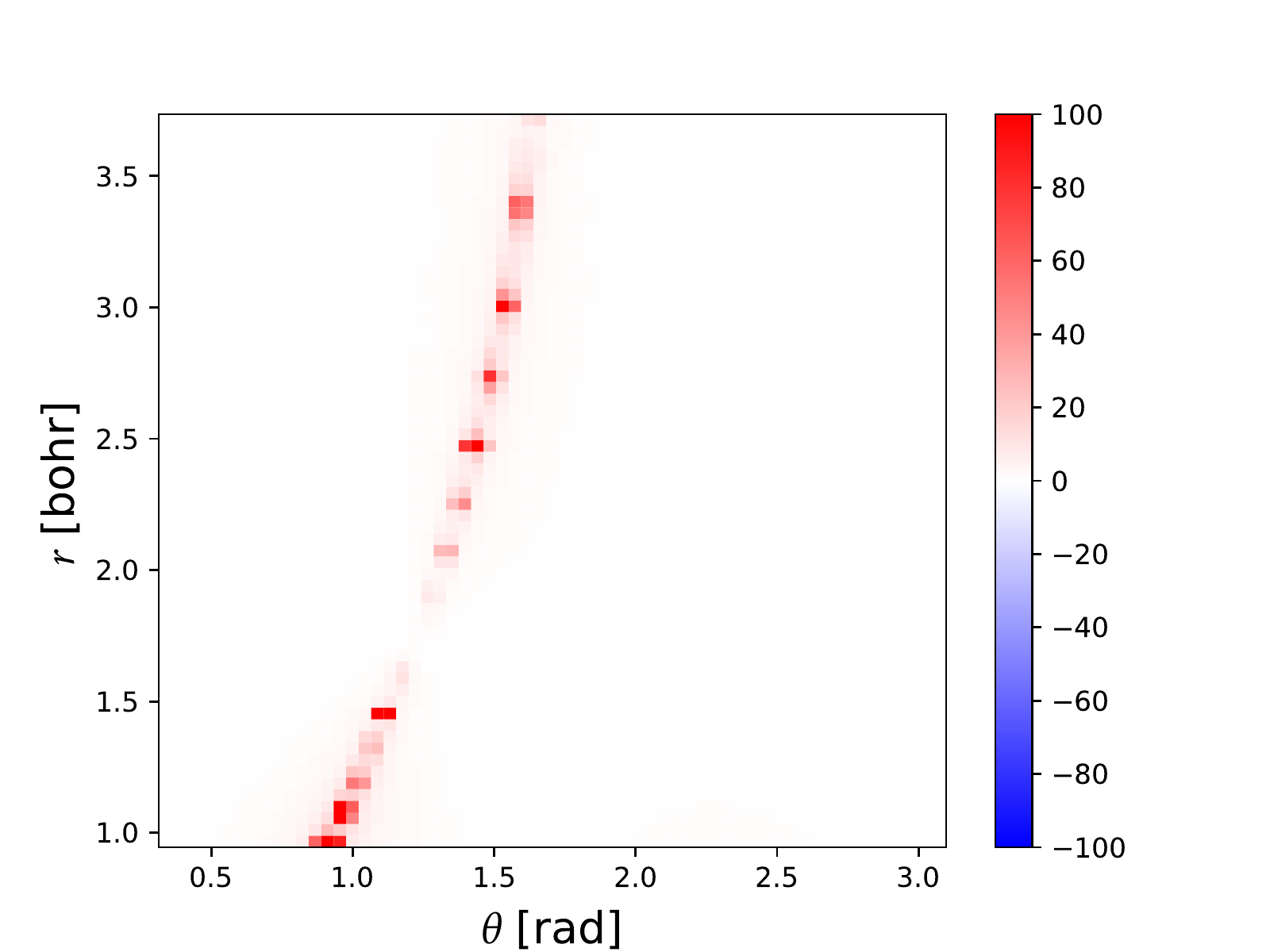}}
\\
\subfloat[NAC vector for the $r$-direction (CASCF)]{\includegraphics[width=6.5cm]{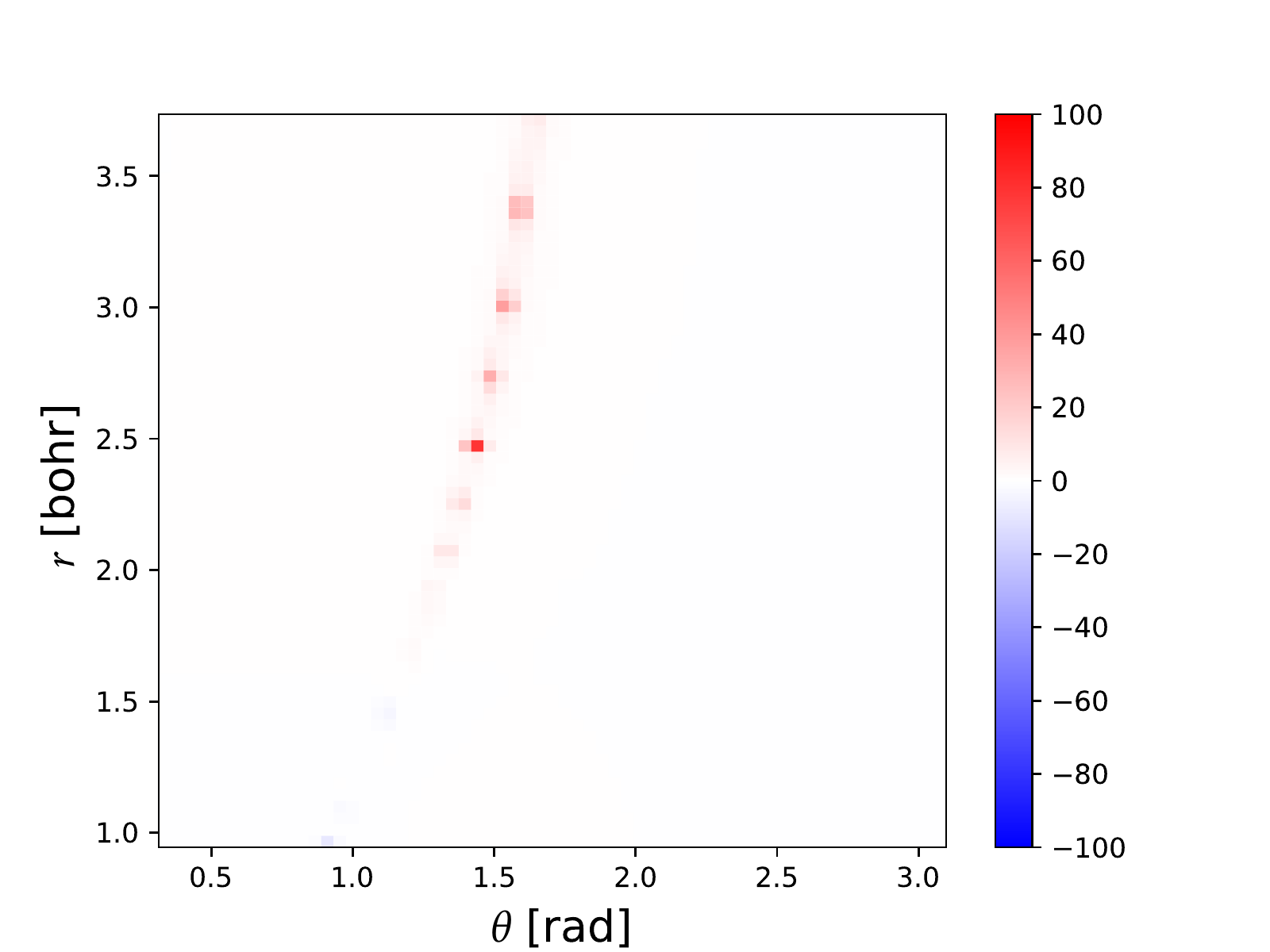}}
\subfloat[NAC vector for the $\theta$-direction (CASSCF)]{\includegraphics[width=6.5cm]{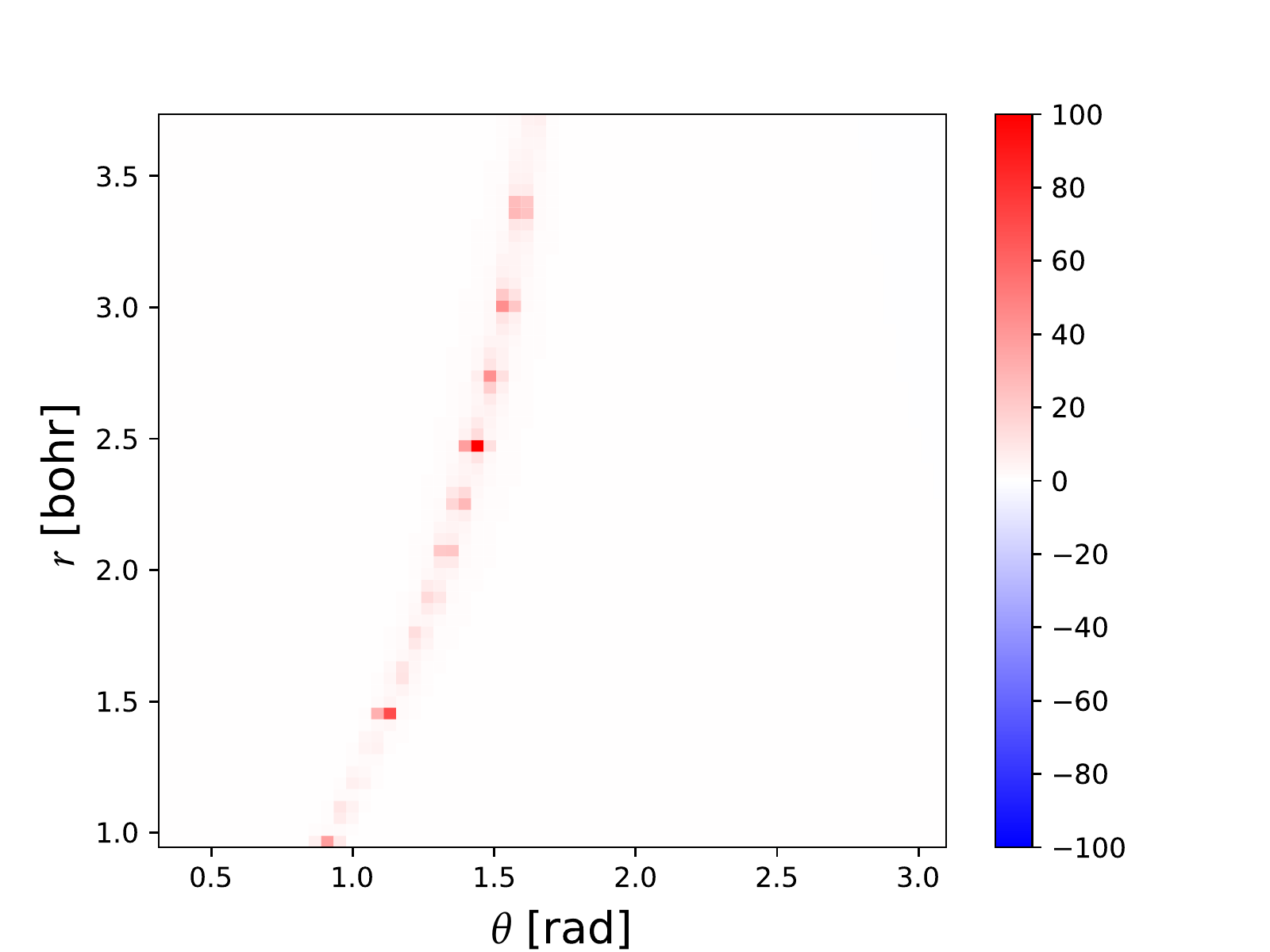}}
\caption{NAC vector's components for the (a)$r$- and (b)$\theta$-directions between the $\tilde{B}^2B_2$ and $\tilde{A}^2A_1$ states for the H$_2$O$^+$ molecule. For comparison, NAC vector's components computed by CASSCF are shown in (c) and (d) for each direction, respectively.}
\label{fig2}
\end{figure}
Figure. \ref{fig2} shows NAC vector components for the $r$- and $\theta$-directions between the $\tilde{B}^2B_2$ and $\tilde{A}^2A_1$ states for the H$_2$O$^+$ molecule.
For comparison, the corresponding components computed by CASSCF are shown.
In the CASSCF computation for NAC vectors, OpenMolcas~\cite{aquilante2020modern} is used and the computational conditions such as the basis set and the active space are set to the same ones as SSVQE computations except the derivative method (the analytical derivative method is used for CASSCF).
Figure. \ref{fig2} shows that there are large NAC vector components on the conical intersection seam between the $\tilde{B}^2B_2$ and $\tilde{A}^2A_1$ states shown in FIG. \ref{fig1} for the both results of SSVQE and CASSCF methods.
There is a slight difference between the NAC vector components of SSVQE and CASSCF, which can be attributed to the error in the finite numerical differentiation of SSVQE.

\subsection{Grid notation of the Hamiltonian}
A non-adiabatic quantum dynamics simulation was conducted on a grid in real space.
The adiabatic PESs and NAC vectors were computed with the method explained in the previous section.
Since these values were computed on the coarse grid, these values should be complemented for a fine grid to conduct the quantum dynamics.
For completion, the three-dimensional spline method is adopted.
The second-order central difference method is used to represent the kinetic energy term,
\begin{equation}
\frac{\partial^2}{\partial r^2} \chi(r, \theta, t) = \frac{\chi_i(r+dr, \theta, t)-2\chi(r, \theta, t)+\chi(r-dr, \theta, t)}{dr^2} + O(dr^2).
\end{equation}
The same applies to the $y$-direction. 
The second-order central difference method is also used for the time derivative,
\begin{equation}
\frac{\partial}{\partial t} \chi(r, \theta, t) = \frac{\chi_i(r, \theta, t+dt)-\chi(r, \theta, t-dt)}{2dt} + O(dt^2).
\end{equation}
Therefore, the time evolution of the system can be computed by the following formula:
\begin{equation}
\chi(r, \theta, t+dt)=\chi(r, \theta, t-dt) - 2dti \hbar H \chi(r, \theta, t),
\end{equation}
where $H$ is the non-adiabatic Hamiltonian with an adiabatic representation under the symmetric extension model and the LHL approximation.
In this study, $dx = 0.0443$ bohr and $dy = 0.0409$ rad were used, and the ranges were set to be 0.9449-3.7352 bohr for $x$ and 0.5236-3.1007 rad for $y$. 
The grid size is then $(N_r, N_\theta) = (64, 64)$.

\section{Results and discussion}

\subsection{Preparation of initial wavepacket}
In this study, a vertical transition was assumed according to the Franck-Condon principle \cite{franck1926elementary}.
This approximation is based on the instantaneous occurrence of electronic transitions compared to the time scale of nuclear motion, so that the ionization occurs without changing the position of the nuclei.
At first, the nuclear wavepacket of the zero-point vibrational state in the electronic ground state PES was computed, and the initial wavepacket was constructed by placing it on the adiabatic PES of the second excited state ($\tilde{B}^2B_2$).
The electron wavepacket in the electronic ground state was obtained by diagonalization of the adiabatic Hamiltonian.
The sum of the zero-point vibrational energies of the $r$ and $\theta$ degrees of freedom known in the experiment should match the energy of this wavepacket.
The energy of the wavepacket and the energies obtained by the photoelectron spectroscopy experiment \cite{reutt1986molecular} are shown in TABLE. \ref{table1}, which 
also shows the results for D$_2$O$^+$.
The results obtained in this study are in good agreement with the experimental values, including the isotopic effect. 
\begin{table}[ht]
  \caption{Comparison of zero-point vibrational energies [hartree].}
  \label{table1}
  \centering
  \begin{tabular}{c|c|c}
    \hline
    System  & This work  & Experiment\cite{reutt1986molecular} \\
    \hline \hline
    H$_2$O$^+$  & 0.0210  & 0.0214 \\
    D$_2$O$^+$  & 0.0149  & 0.0157 \\
    \hline
  \end{tabular}
\end{table}

\subsection{De-excitation process from $\tilde{B}^2B_2$ to $\tilde{A}^2A_1$ in the H$_2$O$^+$ molecule}
\begin{figure}[ht]
\centering
\includegraphics[width=8.6cm]{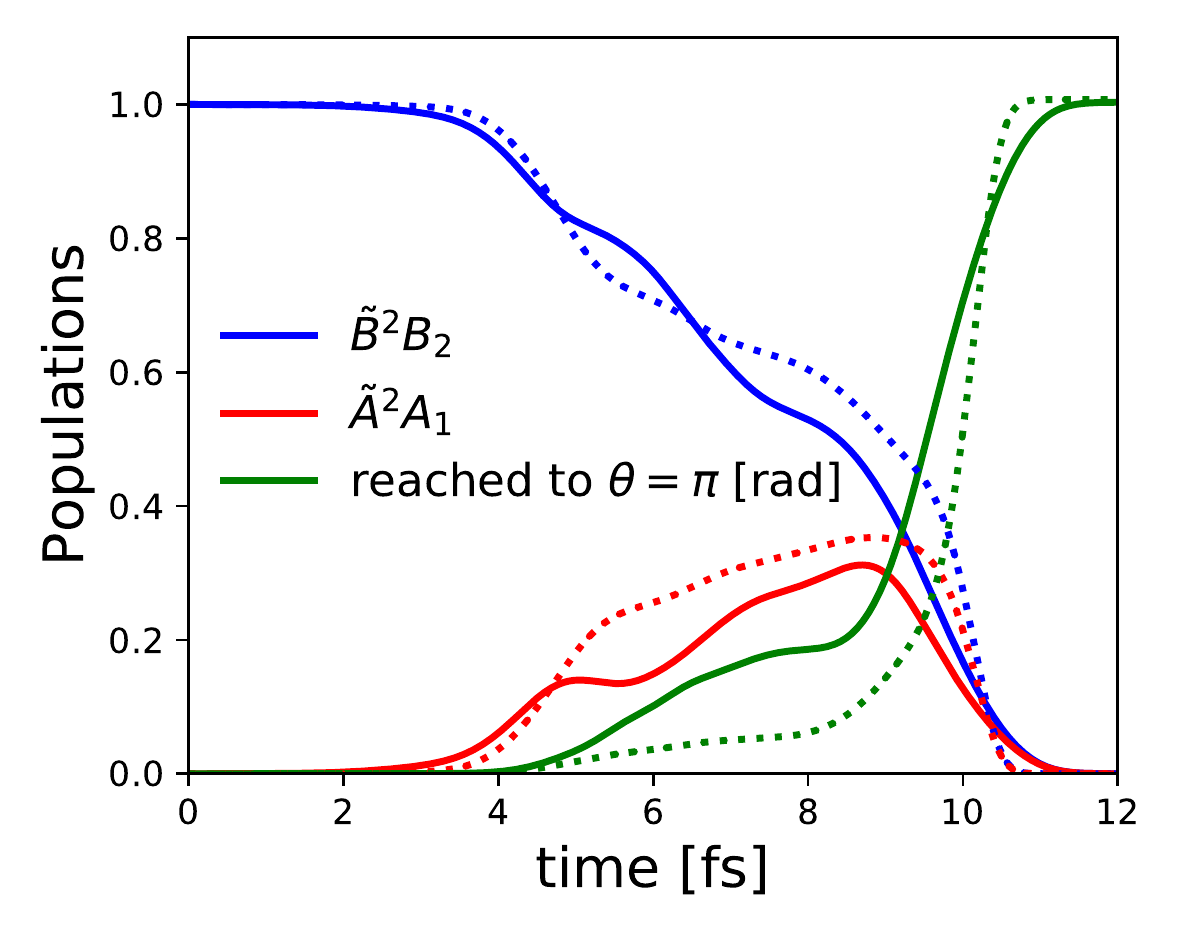}
\caption{Time-evolution of the populations of the $\tilde{B}^2B_2$ and $\tilde{A}^2A_1$ states and the time integrated value of the flux that reached $\theta = \pi$ on $\tilde{A}^2A_1$ for the H$_2$O$^+$ molecule based on the PESs and NAC vectors computed by SSVQE method (solid lines) and CASSCF method (dashed lines).}
\label{fig4}
\end{figure}

\begin{figure}[ht]
\centering
\subfloat[$t=0$ fs on $\tilde{B}^2B_2$]{\includegraphics[width=4.3cm]{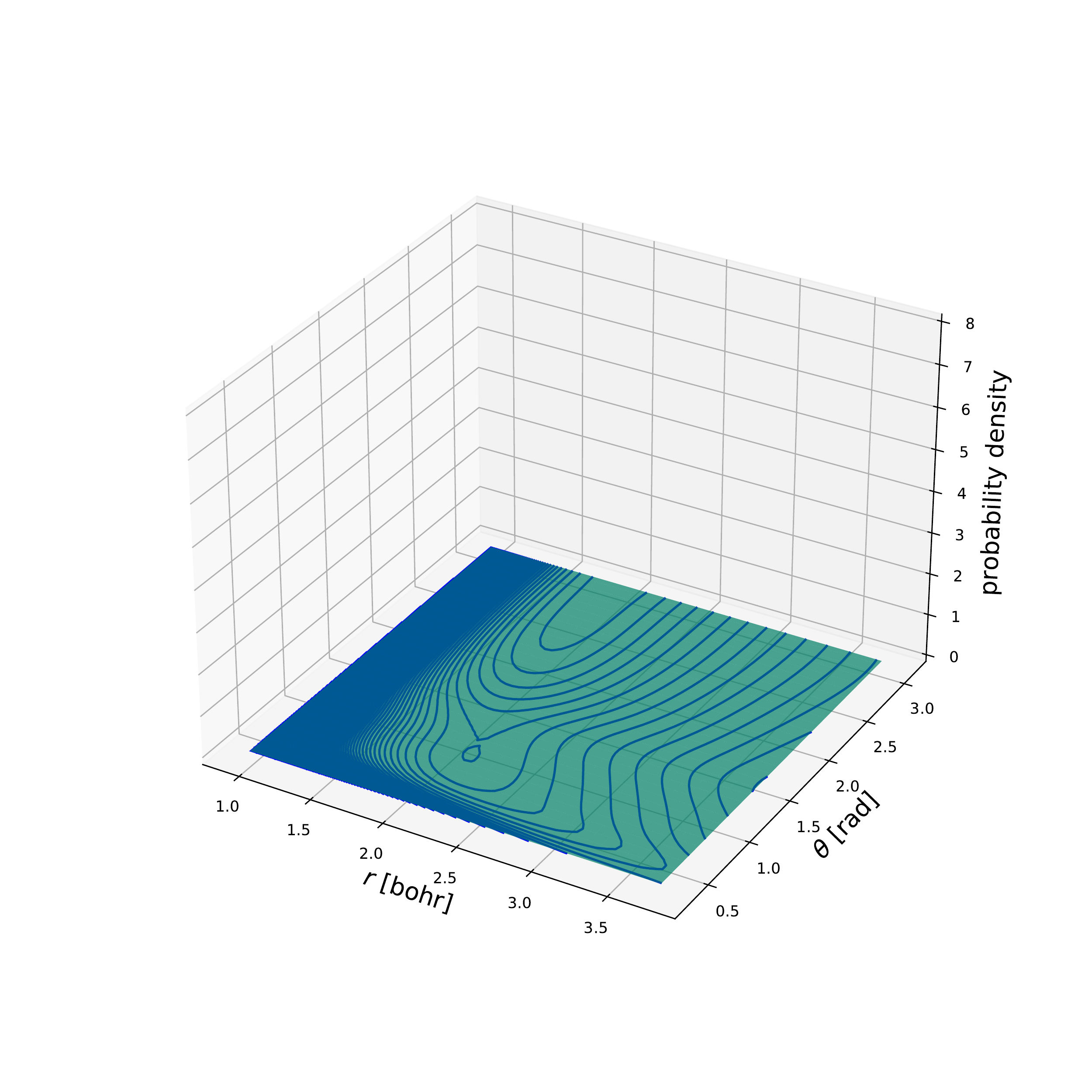}}
\subfloat[$t=0$ fs on $\tilde{A}^2A_1$]{\includegraphics[width=4.3cm]{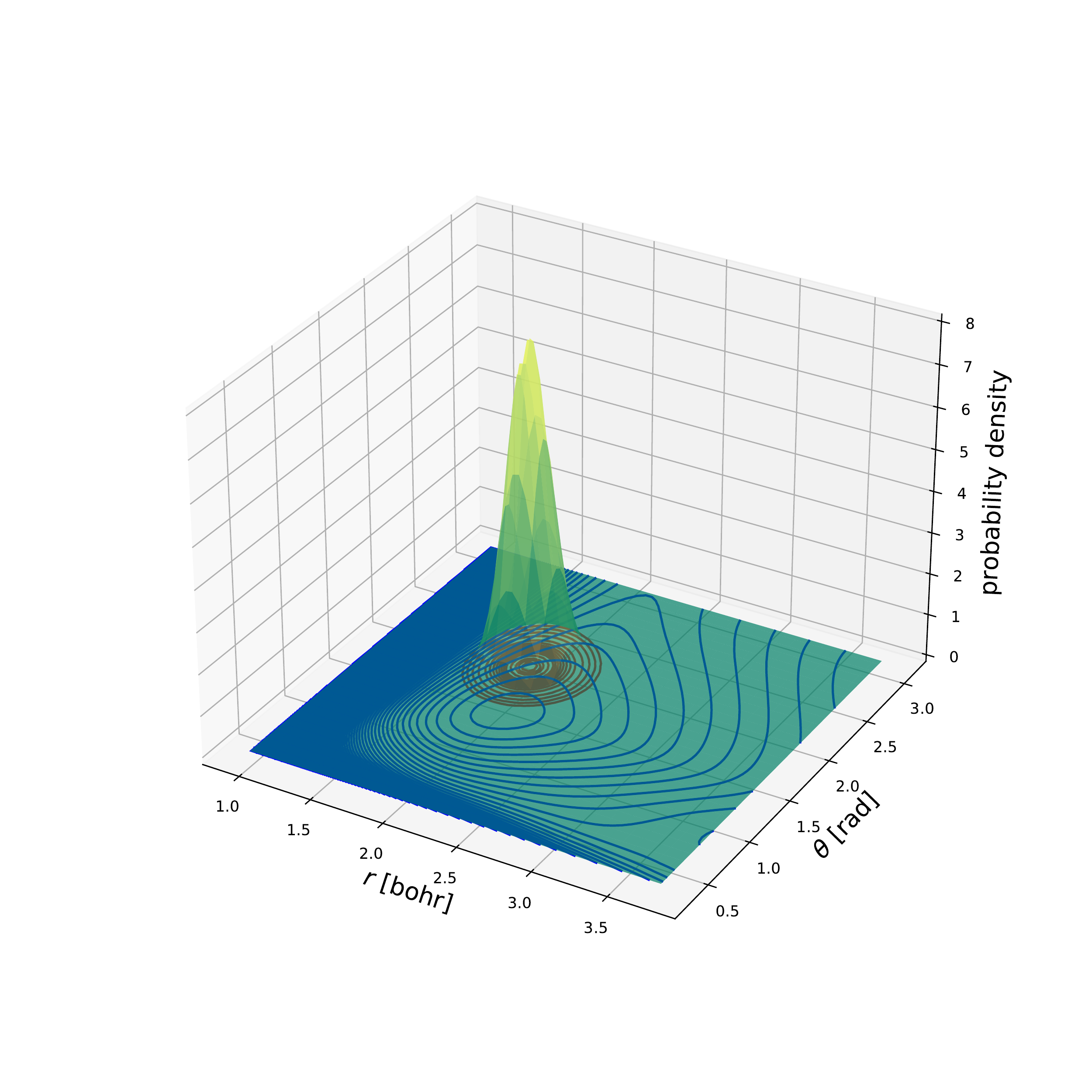}}
\vspace{-10mm}
\subfloat[$t=2.4$ fs on $\tilde{B}^2B_2$]{\includegraphics[width=4.3cm]{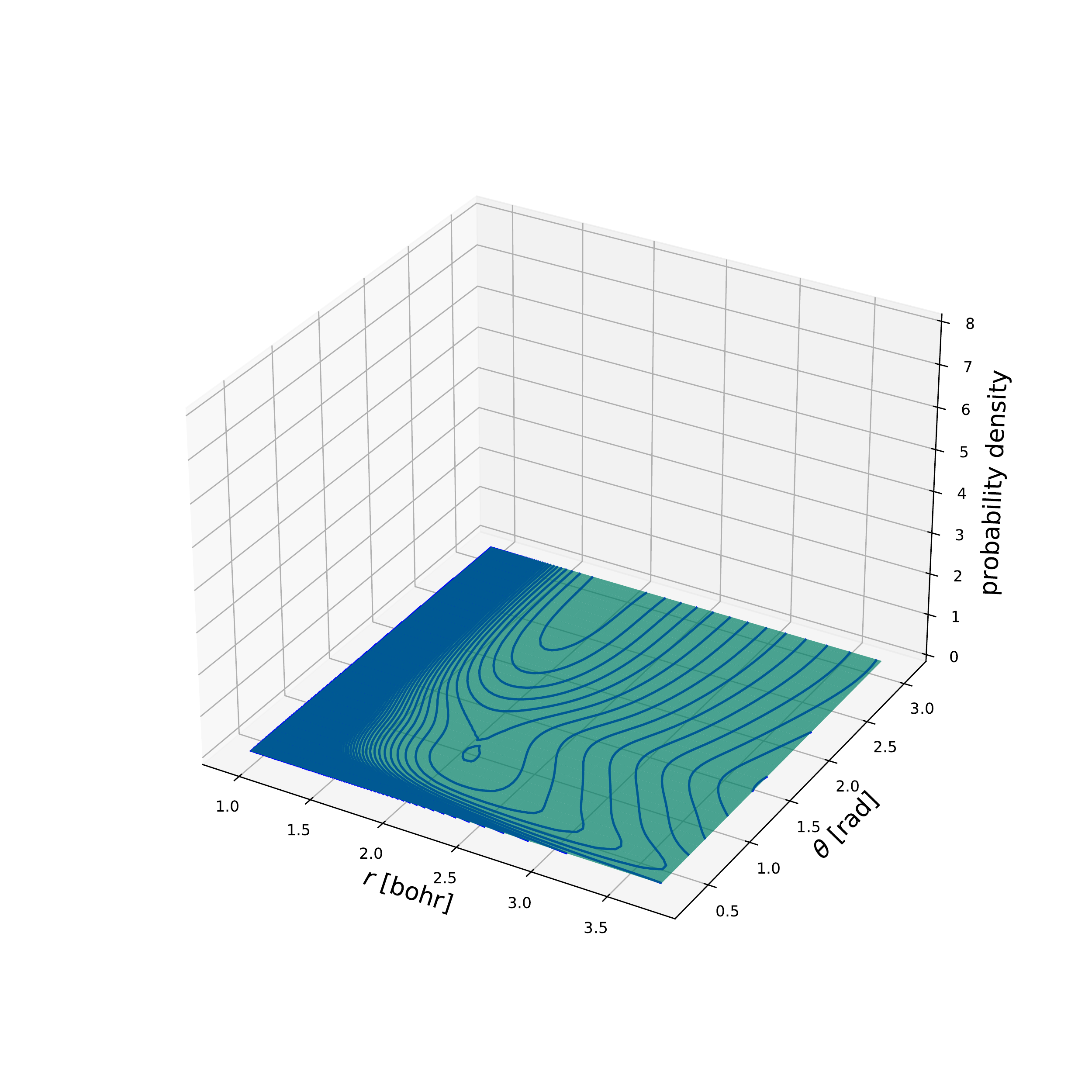}}
\subfloat[$t=2.4$ fs on $\tilde{A}^2A_1$]{\includegraphics[width=4.3cm]{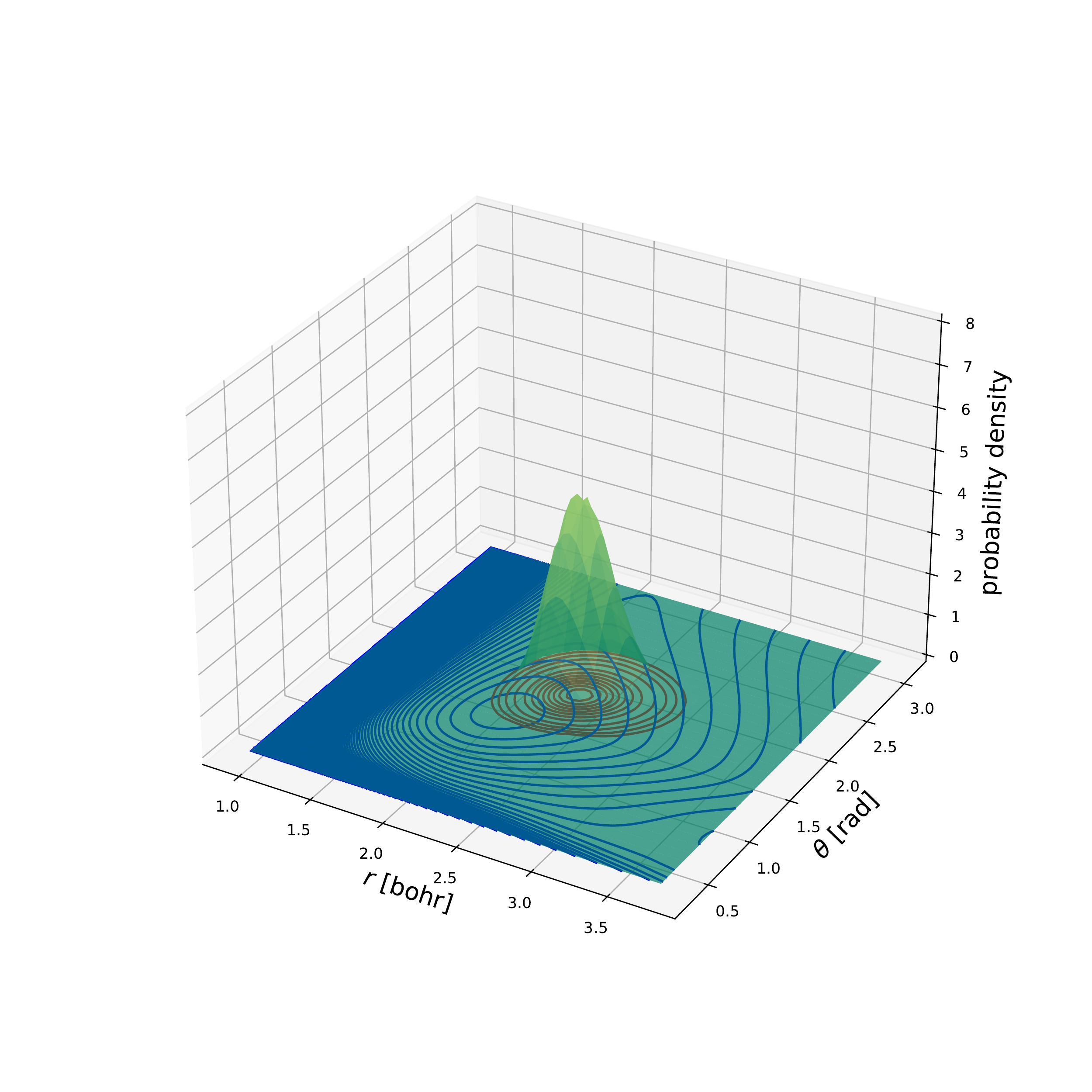}}
\vspace{-10mm}
\subfloat[$t=4.8$ fs on $\tilde{B}^2B_2$]{\includegraphics[width=4.3cm]{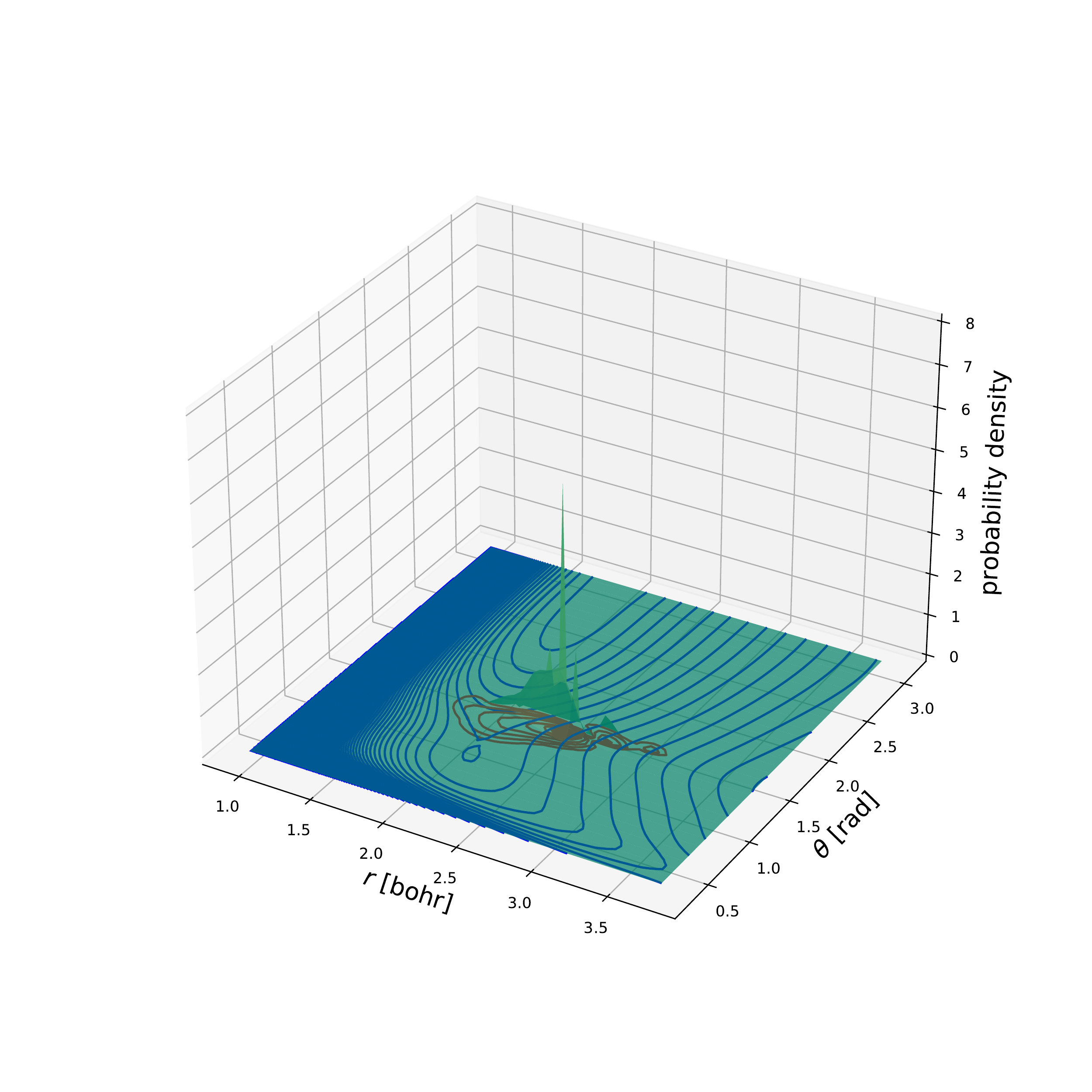}}
\subfloat[$t=4.8$ fs on $\tilde{A}^2A_1$]{\includegraphics[width=4.3cm]{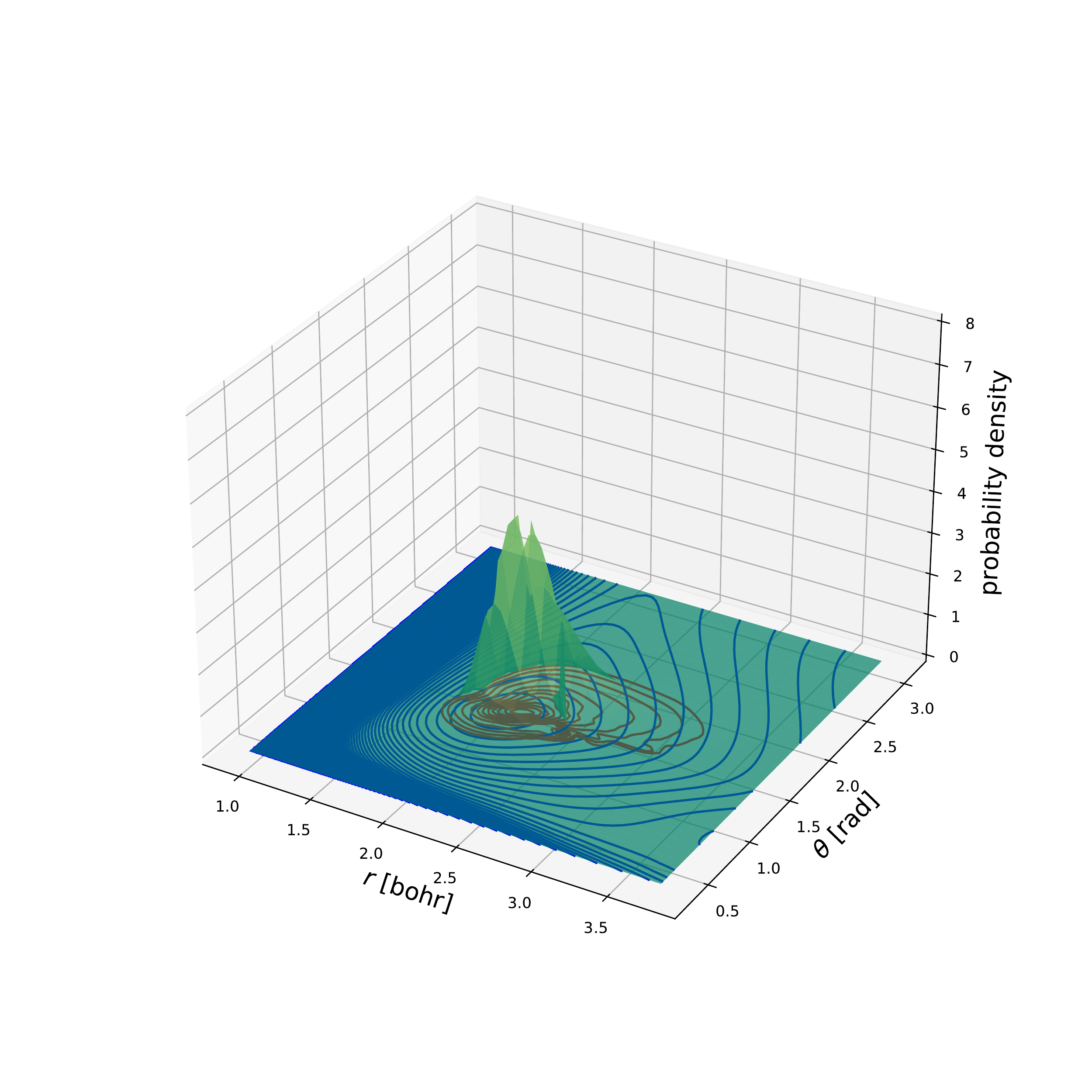}}
\vspace{-10mm}
\subfloat[$t=8.4$ fs on $\tilde{B}^2B_2$]{\includegraphics[width=4.3cm]{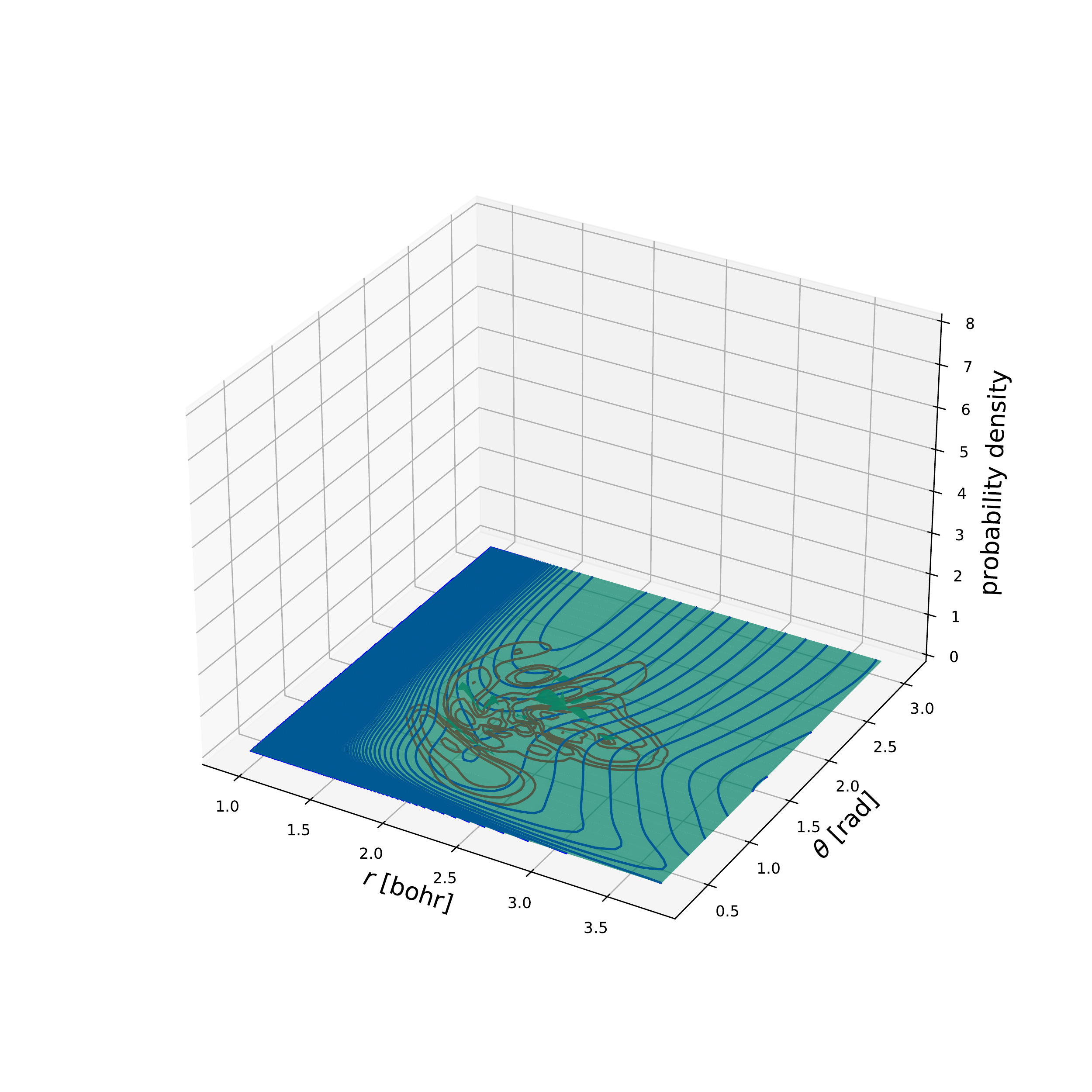}}
\subfloat[$t=8.4$ fs on $\tilde{A}^2A_1$]{\includegraphics[width=4.3cm]{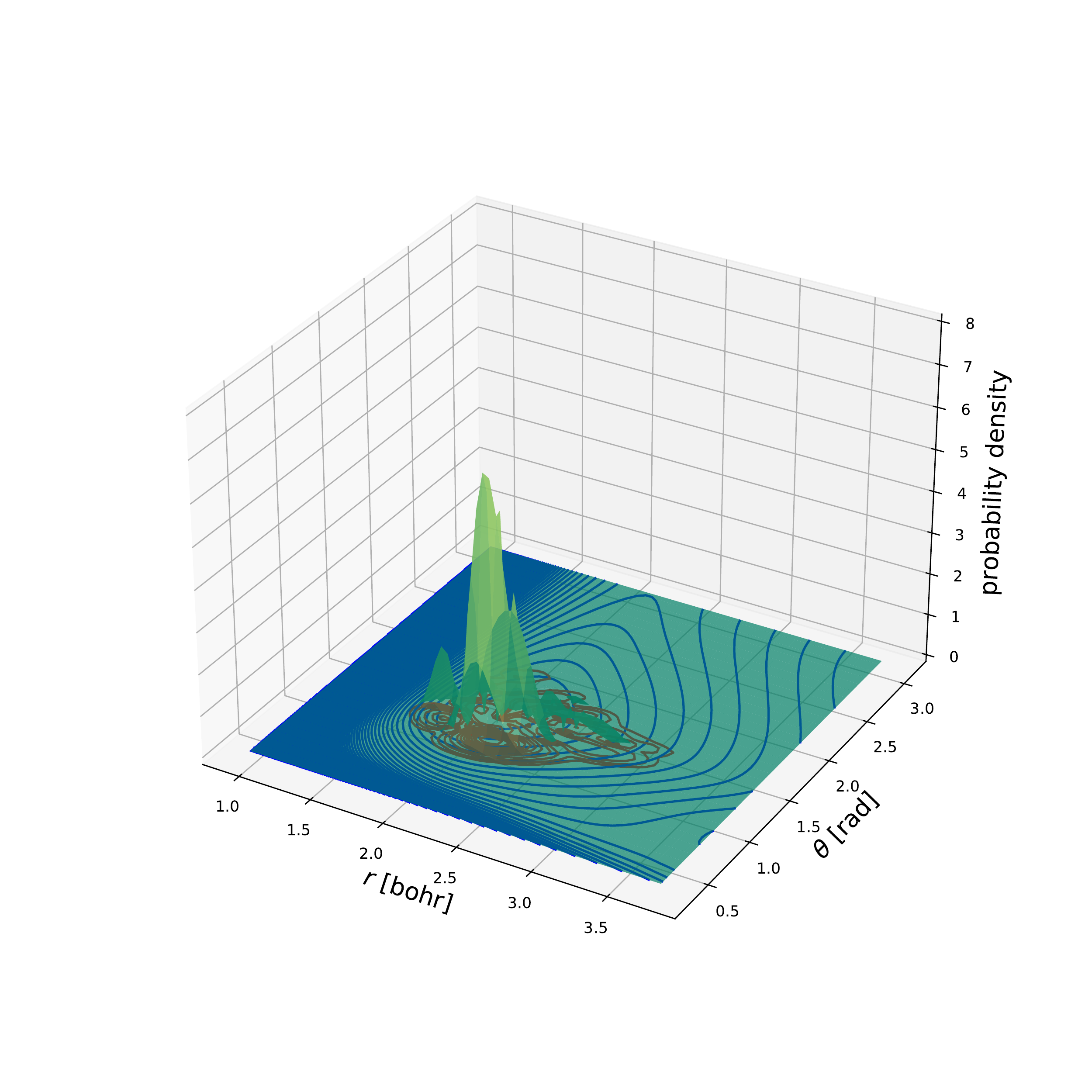}}
\caption{Snapshots of the wavepackets on the $\tilde{B}^2B_2$ and $\tilde{A}^2A_1$ states for H$_2$O$^+$ molecule.}
\label{fig5}
\end{figure}

Figure. \ref{fig4} shows the time-evolution of the populations for the $\tilde{B}^2B_2$ and $\tilde{A}^2A_1$ states.
For comparison, the results simulated with the PESs and NAC vectors computed by CASSCF method are also shown.
The both results are qualitatively consistent, but there is a slight difference in the curve of the population as a function of time, reflecting the difference of NAC vector of SSVQE and CASSCF as shown in Fig. \ref{fig2}.
The transition from $\tilde{B}^2B_2$ to $\tilde{A}^2A_1$ can be confirmed approximately 10 fs after the start of the simulation for the both results.
Figure. \ref{fig5} shows a snapshot of the time-evolution of the wavepacket at each time. 
The wavepacket moves in the $r$-positive direction from $t=0$ to $t=2.4$ fs, then turns around and moves in the $r$-negative direction from $t=2.4$ to $t=4.8$ fs.
The wavepacket reaches the conical intersection near $t=4.8$ fs, and a part of the wavepacket transits from $\tilde{B}^2B_2$ to $\tilde{A}^2A_1$.
However, the NAC vector in the $r$ direction is not so large at this position; therefore, only a part of the transition occurs,
whereas most of the wavepacket transitioned to $\tilde{A}^2A_1$ in the previous study \cite{suarez2015nonadiabatic}. 
The reason for this difference can be considered to be the small active space used to describe the conical intersection, so that the rate of transition from $\tilde{B}^2B_2$ to $\tilde{A}^2A_1$ is smaller than that of the previous report.
The $\theta$ positive motion is subsequently promoted and almost all the wavepackets transition to $\tilde{A}^2A_1$ when passing through the conical intersection.
This is due to the large NAC vector in the $\theta$ direction at this position.
The wavepacket that transitions to $\tilde{A}^2A_1$ has momentum in the $\theta$-positive direction and reaches to $\theta = \pi$, where it is absorbed at the boundary by the absorption potential.
This wavepacket is considered to make the transition to $\tilde{X}^2B_1$ (electronic ground state) by Renner-Teller coupling, which has large coupling near $\theta = \pi$ (the linear geometry).
The motion of the nuclear wavepacket is substantially consistent with the previous reports \cite{suarez2015nonadiabatic,brundle1968high}.

\section{Conclusion}
In this study, we performed a non-adiabatic quantum dynamics simulation in the adiabatic representation using a quantum-classical hybrid scheme.
The non-adiabatic quantum dynamics computations were conducted with the conventional algorithm for classical computers using the Hamiltonian components in the adiabatic representation, which consisted of PESs and NACs computed with NISQ algorithms.
The simulation results for the de-excitation process from $\tilde{B}^2B_2$ (second excited state) to $\tilde{A}^2A_1$ (first excited state) of the H$_2$O$^+$ molecule reproduced the trend in the previous study and consistent with the the results based on the PESs and NAC vectors computed with the CASSCF method, which indicates the effectiveness of this method.
The quantum-classical hybrid calculation scheme proposed in this study may be a promising method for non-adiabatic quantum dynamics simulations in the NISQ era.

\section*{Acknowledgement}
The authors thank Dr. Y. O. Nakagawa for his valuable comments.

\section*{CRediT authorship contribution statement}
\textbf{Hirotoshi Hirai:} Conceptualization, Methodology, Software, Writing - original draft, Investigation, Visualization. \textbf{Sho Koh:} Methodology, Software, Writing - review \& editing, Investigation.

\section*{Declaration of Competing Interest}
The authors declare that they have no known competing financial interests or personal relationships that could have appeared to influence the work reported in this paper.

\clearpage

\bibliographystyle{unsrt}
\bibliography{ref_qc_nonadia_sim2}

\end{document}